\documentclass[twoside]{article}
\usepackage{amsmath}
\usepackage{amssymb}
\usepackage{latexsym}
\usepackage{revsymb}
\usepackage{epsfig}
\usepackage{graphicx}

\newcommand{\bra}[1]{\ensuremath{\left \langle #1 \right |}}
\newcommand{\ket}[1]{\ensuremath{\left | #1 \right \rangle}}
\newcommand{\braket}[2]{\ensuremath{\left \langle #1 \right | \left. #2 \right \rangle}}
\begin{document}

\title{\sf Parts of Quantum States}
\author{\small Nick S. Jones\qquad and\qquad Noah Linden\protect\vspace{-0.09cm}\protect\\
        \small {\tt n.s.jones@bristol.ac.uk}\qquad
               {\tt n.linden@bristol.ac.uk}\protect\vspace{-0.09cm}\protect\\
        \small Department of Mathematics, University of Bristol,\protect\vspace{-0.09cm}\protect\\
        \small University Walk, Bristol BS8 1TW, U.K.}
\date{\small (15th July 2004)}

\maketitle

\begin{abstract}
It is shown that generic $N$-party pure quantum states (with
equidimensional subsystems) are uniquely determined by their
reduced states of just over half the parties; in other words, all
the information in almost all $N$-party pure states is in the set
of reduced states of just over half the parties.  For $N$ even,
the reduced states in fewer than $N/2$ parties are shown to be an
insufficient description of almost all states (similar results
hold when $N$ is odd).  It is noted that Real Algebraic Geometry
is a natural framework for any analysis of parts of quantum
states: two simple polynomials, a quadratic and a cubic, contain
all of their structure. Algorithmic techniques are described which
can provide conditions for sets of reduced states to belong to
pure or mixed states.
\end{abstract}

\maketitle

\section{Introduction}
Given parts of pure, multi-party, quantum states, where parts are
reduced states in subsets of the parties, what does one know about
the whole? One might have expected that the parts leave something
out: that most pure states contain higher-order correlations which
are independent of the lower order ones. This is not the case.

In fact, knowing the reduced states in appropriate subsets of the
parties specifies the state completely \cite{LPW,LW}, the only
state, pure or mixed, consistent with these reduced states is the
original state itself. Thus all the information in most pure
multi-party states resides in these reduced states.

In \cite{LW} upper and lower bounds were found on the size of the
subsets whose reduced states determine the full pure states. It
was shown that reduced states in no more than two-thirds of the
parties are sufficient (generically). The upper bound was
independent of the local dimension, but the lower bound in
\cite{LW} was dimension dependent, varying from about $18.9\%$ for
qubits ($d=2$) and increasing monotonically to half for large
local dimension. A main result of this paper is to improve both
bounds to close to half.

Specifically we show that, considering the $m$-party reduced
states of an $N$-party pure state (each party having
equidimensional Hilbert spaces), a small, interesting (see Fig.
\ref{fig1}), set of reduced states in $m=\lceil N/2\rceil+1$
parties (where $\lceil\;\;\rceil$ and $\lfloor\;\;\rfloor$
indicate rounding up and down to the nearest integer) almost
always form a unique description of the state. We also prove the
lower bound that the $\lfloor N/2\rfloor$-party reduced states of
pure states do not uniquely distinguish them from other pure or
mixed states.

A question, closely related to the above, is, given a set of
reduced states, under what circumstances are they compatible with
\it any \rm states (not necessarily unique or pure) of the full
system. One can think of reduced states as jigsaw puzzle pieces
and the question  is whether there exist one, or more, jigsaw
puzzles of which these are some of the pieces. In \cite{LPW}
examples were given which show this can be a complicated issue and
a number of authors \cite{SHS,Hig,Han,B} have made partial
progress. In Section \ref{jig}, by giving a particular
characterization of quantum states, we describe how Real Algebraic
Geometry (RAG) - the geometric study of real roots of polynomials
\cite{BPR03} - provides a unified mathematical description for
these quantum jigsaw puzzles. Methods from RAG provide systematic
algorithms for tasks like finding  conditions which sets of
reduced states must satisfy in order to be compatible with a
quantum state of the full system. While future developments in
Algorithmic RAG may well make it a useful tool for questions of
this sort, these tasks appear to be hard in a technical sense.
Despite this, given the simplicity of the equations involved, a
sphere and cubic surface (\ref{eta}), there is hope that analytic
progress can be made.

Section \ref{Ia} provides the proof of the upper bound that
reduced states in $\geq\lceil N/2\rceil+1$ parties generically
uniquely characterize the state, Section \ref{Ib} proves the lower
bound. Section \ref{jig} reveals how Real Algebraic Geometry
provides a framework for understanding questions about parts of
quantum states. We conclude with some open questions.
\section{Upper Bound} \label{Ia}

\begin{figure}[h!]
\includegraphics[angle = 0, width = 5.3cm, height = 5.0cm]{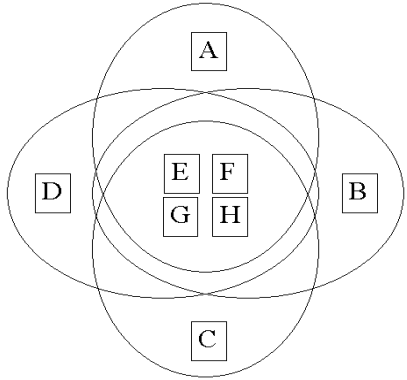}
\caption{\label{fig1}The  four overlapping, five-party reduced
states, $\rho_{AEFGH}$,\ $\rho_{BEFGH},\ \rho_{CEFGH},\
\rho_{DEFGH}$, are almost always a sufficient description of the
eight party pure state $\ket{\phi}_{ABCDEFGH}$ }
\end{figure}
 We consider a pure
quantum state of $N$ parties, each of which has a $d$-dimensional
Hilbert space. In the first instance we will consider the case
that $N$ is even. We will show that almost every such quantum
state is completely determined by its reduced states of $N/2+1$
parties: for almost every pure state of $N$ parties, the unique
state, pure or mixed, consistent with these reduced states is the
original state.

 There are $\begin{pmatrix}
   N \\
   N/2 +1 \\
 \end{pmatrix}$ reduced states of $(N/2 +1)$
parties. In fact, we will show that knowledge of only $N/2$ of
them is sufficient, generically, to uniquely specify an $N$-party
pure state consistent with them. The set of $N/2$ reduced states
may be described as follows: select $N/2$ parties (for convenience
and without loss of generality we take these to be parties
$N/2+1,N/2+2,...,N$), then form $(N/2+1)$-party reduced states by
combining these $N/2$ parties with one further party. Since there
are $N/2$ choices for this last party we have  a set of $N/2$
$(N/2+1)$-party reduced states - this is illustrated in Fig.
\ref{fig1} for $N=8$. We will write the original pure state as:
\begin{equation}\ket{\phi}_N=\sum^d_{i_1...i_N}a_{i_1,i_2,...,i_N}\ket{i_1...i_N}.\end{equation}
Let us consider one of these reduced states obtained by tracing
out parties $2,3,...,N/2$ and call it $\rho^1$:
\begin{eqnarray}\rho^1=\sum^d_{\substack{i_1..i_N,\\j_1,j_{\frac{N}{2}+1}\\..j_N}}
a_{i_1..i_N}a^*_{j_1,i_2..i_{\frac{N}{2}}j_{\frac{N}{2}+1}..j_N}\ket{i_1,i_{\frac{N}{2}+1}..i_N}
\bra{j_1,j_{\frac{N}{2}+1}..j_N}\end{eqnarray}
The most general state of $N$ parties, for which this is a reduced
state, will typically be mixed. To allow for this possibility, we
consider an environment `$E$' with which the system might be
entangled in such a way that the whole system is in a pure state
$\ket{\psi}$. The most general pure state  of system plus
environment, consistent with $\rho^1$, is:
\begin{equation} \ket{\psi^1}=\sum^d_{i_1...i_N}a_{i_1...i_N}\ket{i_1,i_{N/2+1},...i_N}\ket{E^1_{i_2,i_3...i_{N/2}}},\end{equation}  where the states $\ket{E^1_{i_2...i_{N/2}}}$
 are states of parties $2,3$ up to $N/2$, plus the environment. The $d^{N/2-1}$ vectors
$\ket{E^1_{i_2,i_3...i_{N/2}}}$ must satisfy
\begin{equation} \braket{E^1_{i_2...i_{N/2}}}{E^1_{j_2...j_{N/2}}}=\delta_{i_2j_2}...\delta_{i_{N/2}j_{N/2}},\label{ortho}\end{equation}
to ensure that the state $\ket{\psi^1}$,  when reduced to parties
$(1,N/2+1,...,N)$, yields $\rho^1$. In fact we will not need to
use  equations (\ref{ortho}) to prove our result that the
$(N/2+1)$-party reduced states uniquely specify a pure quantum
state of $N$ parties.

It will be convenient to rewrite each state
$\ket{E^1_{i_2,i_3...i_{N/2}}}$ explicitly as an entangled state
of system plus environment: \begin{equation}
\ket{E^1_{i_2,i_3...i_{N/2}}}=\sum^d_{j_2...j_{N/2}}\ket{j_2...j_{N/2}}\ket{e^1_{i_2...i_{N/2};j_2...j_{N/2}}},\end{equation}
where $\ket{e^1_{i_2...i_{N/2};j_2...j_{N/2}}}$ are states of the
environment which need not be normalized or orthogonal, but must
only be such that (\ref{ortho}) holds. The upper index on the
vectors $\ket{e}$ denotes the fact that these are the states of
the environment arising from purifications of $\rho^1$.

$\ket{\psi^1}$, the most general pure state of system and
environment consistent with $\rho^1$, is then:
\begin{equation}
\ket{\psi^1}=\sum^d_{\substack{i_1..i_N,\\j_2..j_{\frac{N}{2}}}}a_{\substack{i_1,j_2..j_{\frac{N}{2}},i_{\frac{N}{2}+1}...i_N}}\ket{i_1..i_N}\ket{e^1_{j_2..j_{\frac{N}{2}};i_2..i_{\frac{N}{2}}}}.
\label{psi1}\end{equation} If  $\rho^2$ is the reduced state of
parties $(2,N/2+1, N/2+2,...,N)$ arising from the original state
$\ket{\phi}_N$ then $\ket{\psi^2}$, the most general state of
system plus environment consistent with $\rho^2$, is:
\begin{eqnarray}
\ket{\psi^2}=\sum^d_{\substack{i_1..i_N,\\j_1,j_3..\\..j_{\frac{N}{2}}}}a_{\substack{j_1,i_2,j_3..j_{\frac{N}{2}},i_{\frac{N}{2}+1}..i_N}}\ket{i_1..i_N}\ket{e^2_{j_1,j_3..j_{\frac{N}{2}};i_1,i_3..i_{\frac{N}{2}}}}.
\label{psi2}\end{eqnarray} $\rho^1$ and $\rho^2$ are both
$(N/2+1)$-party reduced states of the original pure state
$\ket{\phi}_N$. $\rho^1$ and $\rho^2$ have parties $N/2...N$ in
common: $\rho^1$ is a state of party $1$ and parties $(N/2+1)...N$
and $\rho^2$, party $2$ and parties $(N/2+1)...N$. There are
analogous states $\ket{\psi^3}...\ket{\psi^{N/2}}$ arising from
the reduced states $\rho^3...\rho^{N/2}$. For there to exist one
or more pure states which have $\rho^1...\rho^{N/2}$ as their
reduced states, these states $\ket{\psi^1},...,\ket{\psi^{N/2}}$
must be equal.

We will show below that this requirement leads to:
\begin{equation}
\ket{e^1_{j_2...j_{N/2};i_2...i_{N/2}}}=\delta_{j_2i_2}\delta_{j_3i_3}...\delta_{j_{N/2}i_{N/2}}\ket{e^1_{11...1;11...1}},
\label{delta}\end{equation} all environment states are  multiples
of a fixed state $\ket{e^1_{11...1;11...1}}$ so that
$\ket{\psi^1}=\ket{\phi}_N\ket{e^1_{11...1;11...1}}$.
$\ket{\psi^1}$ is a product state of the original pure state with
a state of the environment. Hence, the unique state of the system
consistent with the reduced states is $\ket{\phi}_N$, the original
state.

We now prove this result. We first deduce conditions on the states
of the environment imposed by  requiring
$\ket{\psi^1}=\ket{\psi^2}$. From (\ref{psi1},\ref{psi2}) and
comparing terms:
\begin{eqnarray}
\sum^d_{j_2...j_{\frac{N}{2}}}a_{i_1,j_2..j_{\frac{N}{2}},i_{\frac{N}{2}+1}..i_N}\ket{e^1_{j_2..j_{\frac{N}{2}};i_2..i_{\frac{N}{2}}}}=
\sum^d_{j_1,j_3...j_{\frac{N}{2}}}a_{j_1,i_2,j_3..j_{\frac{N}{2}},i_{\frac{N}{2}+1}..i_N}\ket{e^2_{j_1,j_3..j_{\frac{N}{2}};i_1,i_3..i_{\frac{N}{2}}}},\label{at}\end{eqnarray}
for all $(i_1...i_N)$.  Writing $\mathbf{u}=
i_{N/2+1}...i_N,\;\;\mathbf{v}=
i_3...i_{N/2},\;\;\mathbf{v'}=j_3...j_{N/2}$ the above becomes:
\begin{equation}
\sum_{j_2,\mathbf{v'}}a_{i_1,j_2,\mathbf{v'},\mathbf{u}}\ket{e^1_{j_2,\mathbf{v'};i_2,\mathbf{v}}}=\sum_{j_1,\mathbf{v'}}a_{j_1,i_2,\mathbf{v'},\mathbf{u}}\ket{e^2_{j_1,\mathbf{v'};i_1,\mathbf{v}}},
\end{equation}
for all $(i_1,i_2,\mathbf{u},\mathbf{v})$. It is convenient to
rewrite this as:
\begin{eqnarray}
& &\sum_{\substack{j_2\neq
i_2,\mathbf{v'}}}a_{i_1,j_2,\mathbf{v'},\mathbf{u}}\ket{e^1_{j_2,\mathbf{v'};i_2,\mathbf{v}}}-\sum_{j_1\neq
i_1,\mathbf{v'}}a_{j_1,i_2,\mathbf{v'},\mathbf{u}}\ket{e^2_{j_1,\mathbf{v'};i_1,\mathbf{v}}}\nonumber\\
& &\qquad\qquad
+\sum_{\mathbf{v'}}a_{i_1,i_2,\mathbf{v'},\mathbf{u}}
\Big[\ket{e^1_{i_2,\mathbf{v'};i_2,\mathbf{v}}}-\ket{e^2_{i_1,\mathbf{v'},i_1,\mathbf{v}}}\Big]=0
\end{eqnarray}
for all $(i_1,i_2,\mathbf{u},\mathbf{v})$. Note that the index
$\mathbf{u}$ takes $d^{N/2}$ values. Let us set
$(i_1,i_2,\mathbf{v})=(c_1,c_2,c_{\mathbf{v}})$ for some constant
integers $c_1,c_2,c_{\mathbf{v}}$. We will consider sets of
$d^{N/2}$ equations in which $\mathbf{u}$ varies and the other
indices are fixed. The above equation becomes:
\begin{eqnarray}
& &\sum_{\substack{j_2\neq
c_2\\\mathbf{v'}}}a_{c_1,j_2,\mathbf{v'},\mathbf{u}}\ket{e^1_{j_2,\mathbf{v'};c_2,c_{\mathbf{v}}}}-\sum_{\substack{j_1\neq
c_1\\\mathbf{v'}}}a_{j_1,c_2,\mathbf{v'},\mathbf{u}}\ket{e^2_{j_1,\mathbf{v'};c_1,c_{\mathbf{v}}}}\nonumber\\
& &\qquad
\qquad+\sum_{\mathbf{v'}}a_{c_1,c_2,\mathbf{v'},\mathbf{u}}
\Big[\ket{e^1_{c_2,\mathbf{v'};c_2,c_{\mathbf{v}}}}-\ket{e^2_{c_1,\mathbf{v'},c_1,c_{\mathbf{v}}}}\Big]=0
.\end{eqnarray} For a given $\mathbf{u}$, the terms in
$\ket{e^1_{j_2,\mathbf{v'};c_2,c_{\mathbf{v}}}}$, $(j_2 \neq
c_2)$, $\ket{e^2_{j_1,\mathbf{v'};c_1,c_{\mathbf{v}}}}$, $(j_1\neq
c_1)$ and
$\Big[\ket{e^1_{c_2,\mathbf{v'};c_2,c_{\mathbf{v}}}}-\ket{e^2_{c_1,\mathbf{v'},c_1,c_{\mathbf{v}}}}\Big]$
all have different coefficients for all $j_1,j_2,\mathbf{v'}$
values. Similarly on comparing any two equations with different
$\mathbf{u}$ values one notes that the coefficients of
$\ket{e^1_{j_2,\mathbf{v'};c_2,c_{\mathbf{v}}}},\ket{e^2_{j_1,\mathbf{v'};c_1,c_{\mathbf{v}}}}$
and\hfil\break
$\Big[\ket{e^1_{c_2,\mathbf{v'};c_2,c_{\mathbf{v}}}}-\ket{e^2_{c_1,\mathbf{v'},c_1,c_{\mathbf{v}}}}\Big]$
are all indexed by $\mathbf{u}$ - so between equations with
different $\mathbf{u}$ values, the coefficients will differ. Let
us treat
$\ket{e^1_{j_2,\mathbf{v'};c_2,c_{\mathbf{v}}}},\ket{e^2_{j_1,\mathbf{v'};c_1,c_{\mathbf{v}}}}$
and
$\Big[\ket{e^1_{c_2,\mathbf{v'};c_2,c_{\mathbf{v}}}}-\ket{e^2_{c_1,\mathbf{v'},c_1,c_{\mathbf{v}}}}\Big]$
for all $\mathbf{v'}, j_1\neq c_1,j_2\neq c_2$ as the variables in
a set of $d^{N/2}$ homogeneous equations indexed by $\mathbf{u}$.
There are $2d^{N/2-1}-d^{N/2-2}$ of these variables (recall that
$(c_1,c_2,c_{\mathbf{v}})$ are fixed). Since we know that all of
the coefficients of the variables are distinct, and generically
there is no relationship between them, we can pick a subset of
$2d^{N/2-1}-d^{N/2-2}$ equations from the $d^{N/2}$ and solve to
show $\ket{e^1_{j_2,\mathbf{v'};c_2,c_{\mathbf{v}}}}=0$ for all
$\mathbf{v'},j_2\neq c_2$,
$\ket{e^2_{j_1,\mathbf{v'};c_1,c_{\mathbf{v}}}}=0$ for all
$\mathbf{v'},j_1\neq c_1$ and
$\ket{e^1_{c_2,\mathbf{v'};c_2,c_{\mathbf{v}}}}=\ket{e^2_{c_1,\mathbf{v'},c_1,c_{\mathbf{v}}}}$
for all $\mathbf{v'}$. By repeating the above procedure for
different $c_2$ and $c_{\mathbf{v}}$ values we find
$\ket{e^1_{j_2,\mathbf{v'};i_2,\mathbf{v}}}=\delta_{j_2i_2}\ket{e^1_{1,\mathbf{v'};1,\mathbf{v}}}$.

Having considered $\ket{\psi^1}=\ket{\psi^2}$ we now study
$\ket{\psi^1}=\ket{\psi^r}$ for all $r$, $1<r \leq N/2$, and find
similarly
$\ket{e^1_{j_r,\mathbf{w'};i_r,\mathbf{w}}}=\delta_{j_ri_r}\ket{e^1_{1,\mathbf{w'};1,\mathbf{w}}}$
where $\mathbf{w}=i_2,i_3...i_{r-1},i_{r+1}...i_{N/2}$ and
$\mathbf{w'}=j_2,j_3...j_{r-1},j_{r+1}...j_{N/2}$. Combining these
results we obtain (\ref{delta}). A similar analysis can be
repeated for $N$ odd. Here a set of $(N-1)/2$ reduced states in
$(N+3)/2$ parties, generically, uniquely determines the quantum
state.

\section{Lower Bound} \label{Ib}

We now derive a lower bound: the reduced states of this fraction
of the parties are not sufficient (generically) to allow one to
reconstruct a unique state of the full system. Again one starts
with a pure state $\ket{\phi}_N$ of $N$ parties. We calculate all
$m$-party reduced states. We then ask whether these
$\begin{pmatrix}
   N \\
   m \\
 \end{pmatrix}$ $m$-party reduced states are consistent with a unique starting
state (i.e. $\ket{\phi}$) or whether there are other (typically
mixed) states of $N$-parties consistent with these
$\begin{pmatrix}
   N \\
   m \\
 \end{pmatrix}$
$m$-party reduced states. We show that reduced states of $\lfloor
N/2\rfloor$ parties do not contain enough information to
reconstruct a unique state.

Consider  a starting  $N$-party pure state $\ket{\phi}_N$. Let us
calculate from it the reduced state $\rho^{(1)}$ of the first $m$
parties ($m\leq N/2$). This will typically have rank $d^m$, and we
may expand it as
\begin{equation}
\rho^{(1)}=\sum_{a=1}^{d^m}\ket{v^{(1)}_a}\bra{v^{(1)}_a},
\end{equation}
where $\ket{v^{(1)}_a}$ are non-normalised eigenvectors of
$\rho^{(1)}$.
 The
most general state of $N$ parties consistent with this reduced
state will be mixed and of rank $d^N$. It will be convenient to
purify this mixed state of $N$ parties to a state
$\ket{\psi^{(1)}}$ of $N$ parties plus an environment of dimension
$d^N$. So,
\begin{eqnarray}
&&\ket{\psi^{(1)}}=\sum_{a=1}^{d^m}\ket{v^{(1)}_a}\ket{F^{(1)}_a},\\
&&\ket{F^{(1)}_a}=\sum_{i_{m+1}...i_N}^d\ket{i_{m+1}...i_N}\ket{f^{(1)}_{a i_{m+1}...i_N}},\\
&&\ket{\psi^{(1)}}=\sum_{a=1}^{d^m}\sum_{i_{m+1}...i_N}^d\ket{v^{(1)}_a}\ket{i_{m+1}...i_N}\ket{f^{(1)}_{a
i_{m+1}...i_N}}.
\end{eqnarray}

The $d^m\times d^{N-m}=d^N$ states $\ket{f^{(1)}_{a
i_{m+1}...i_N}}$ of the environment must satisfy:
\begin{equation}
\sum_{i_{m+1}...i_N}\braket{f^{(1)}_{a
i_{m+1}...i_N}}{f^{(1)}_{bi_{m+1}...i_N}}=\delta_{ab},\label{tapas}
\end{equation}
for $\ket{\psi^{(1)}}$ to reduce to $\rho^{(1)}$, the original
reduced state of the first $m$-parties. In the following we will
group indices, writing $\ket{f^{(1)}_{a
i_{m+1}...i_N}}=\ket{f^{(1)}_{\mu}}$, $\mu=1,...,d^N$. Given a
particular purification of a mixed $m$-party state one can form
others by a unitary transformation on the environment alone. Let
us fix this freedom by rotating the states $ \ket{f^{(1)}_{\mu}}$
so that they are expressed in the following way with respect to a
fixed orthonormal basis, $\ket{1},\ket{2},...,\ket{d^N}$, of the
environment:
\begin{eqnarray}
&&\ket{f^{(1)}_{1}}=\alpha^{(1)}_{1,1}\ket{1}\nonumber\\
&&\ket{f^{(1)}_{2}}=\alpha^{(1)}_{2,1}\ket{1}+\alpha^{(1)}_{2,2}\ket{2}\nonumber\\
&&\ket{f^{(1)}_{3}}=\alpha^{(1)}_{3,1}\ket{1}+\alpha^{(1)}_{3,2}\ket{2}+\alpha^{(1)}_{3,3}\ket{3}\nonumber\\
&&\vdots \nonumber\\
&&\ket{f^{(1)}_{d^{N}}}=\alpha^{(1)}_{d^N,1}\ket{1}+\alpha^{(1)}_{d^N,2}\ket{2}+..+\alpha^{(1)}_{d^{N},d^{N}}\ket{d^N}
\end{eqnarray}

We may use the unitary freedom on the environment to arrange for
the $\alpha^{(1)}_{\mu,\mu}$ to be real; the remaining
coefficients $\alpha^{(1)}_{\mu,\nu}$ ($\nu<\mu$) will be complex.
The total number of real parameters in the states  $
\ket{f^{(1)}_{\mu}}$ is $d^N\times d^N=d^{2N}$.

Consider a different set of $m$ parties. Again calculate the
reduced states of $\ket{\phi}_N$ for these parties and purify it
to a pure state  $ \ket{\psi^{(2)}}$ of the system plus
environment using the environment spanned by
$\ket{1},\ket{2},...,\ket{d^N}$. This will lead to a different set
of $d^N$  states of the environment, $ \ket{f^{(2)}_{\mu}}$ . We
will shortly be requiring that $ \ket{\psi^{(1)}}=
\ket{\psi^{(2)}}$ and therefore the states  $ \ket{f^{(2)}_{\mu}}$
lie in the span of the fixed basis of
$\ket{1},\ket{2},...,\ket{d^N}$. The number of real parameters
describing the $d^N$ vectors $ \ket{f^{(2)}_{\mu}}$ is $2d^{2N}$.
Proceeding in this way for each possible set of $m$ parties, we
find that the total number of parameters describing the $
\ket{f^{(A)}_{\mu}}$ (for $A=1,...,\begin{pmatrix}
   N \\
   m \\
 \end{pmatrix}$) is:
\begin{equation}
P=d^{2N} + 2\bigg[\begin{pmatrix}
   N \\
   m \\
 \end{pmatrix} -1\bigg]d^{2N}=d^{2N}\bigg[
2\begin{pmatrix}
   N \\
   m \\
 \end{pmatrix} -1\bigg]
\end{equation}

\subsection*{ Counting constraint equations}

There are two types of constraint on the $\ket{f^{(A)}_{\mu}}$.
Firstly each set of $\ket{f^{(A)}_{\mu}}$, for fixed $A$, must
satisfy equations like (\ref{tapas}) which ensure that the
purifications associated with these $\ket{f^{(A)}_{\mu}}$ reduce
to the correct $m$-party reduced state.

The number of constraint equations for each $A$ is $d^{2m}$, so
that the total number of equations of this type is
$\begin{pmatrix}
   N \\
   m \\
 \end{pmatrix}d^{2m}$ (this is an overestimate since many of these equations
will be dependent).

The second set of constraints are that the purifications arising
from each set of $m$ parties are equal. The fact that
$\ket{\psi^{(1)}}= \ket{\psi^{(2)}}$ leads to $2d^{2N}$ equations.
Thus the total number of equations of this type (equating
$\ket{\psi^{(1)}}$ to each of the other purifications) is
$2d^{2N}(\begin{pmatrix}
   N \\
   m \\
 \end{pmatrix}-1)$.

It may be that not all the constraints described are independent.
An upper bound on the number of independent constraints is thus:
\begin{equation}
C=2d^{2N}\bigg[\begin{pmatrix}
   N \\
   m \\
 \end{pmatrix}-1\bigg]+\begin{pmatrix}
   N \\
   m \\
 \end{pmatrix}d^{2m}.
\end{equation}

\subsection*{Finding the lower bound on m}

From above:
\begin{equation}
 P-C=d^{2N}-\begin{pmatrix}
   N \\
   m \\
 \end{pmatrix}d^{2m}.\end{equation} Thus for $m\leq N/2$, $d\geq2$, $P>C$. The
number of constraints is insufficient to uniquely specify the
parameters and so, for $N$ even, the reduced states in $N/2$
parties do not uniquely determine the state; for $N$ odd,  reduced
states in $(N-1)/2$ parties do not uniquely define the state.

\section{Quantum jigsaw puzzles} \label{jig}
The considerations in previous sections lead us to address the
following general (and related) questions. First, \it{given} \rm
some partial information about a putative quantum state (for
example a set of reduced states) how does one check that this
partial information is indeed consistent with one (or more) pure
or mixed states. For example, for a state of three parties $A,B,C$
one might be given three particular two-party density matrices
$\sigma^{AB},\tau^{BC},\eta^{AC}$. One would like to be able to
determine whether these three states are possible reduced states
of any three party quantum state $\rho^{ABC}$.

A second, more general, question is to determine conditions under
which partial information about a putative quantum state is in
fact legitimate. Referring to the example in the previous
paragraph, rather than aiming to produce an algorithm which
determines whether the \em{given} \rm
$\sigma^{AB},\tau^{BC},\eta^{AC}$ are reduced states of any
$\rho^{ABC}$, one would like to have a set of conditions satisfied
by the two-party reduced states of $\rho^{ABC}$ (pure or mixed).
Despite the similarity of these two questions it should be clear
that algorithms for solving them may be rather different.

These questions are two version of the task of, given a set of
parts of quantum states, testing if there  exists any state with
these parts. In the same spirit, one might take a jumble of puzzle
pieces and check to see if there exists a jigsaw puzzle of which
these are the parts. One could test if a set of pieces make a
whole puzzle by either trying to make that puzzle (testing a given
set of parts) or checking the pieces against a system of rules
(finding general compatibility conditions).

In this section we aim to show that both questions fall within the
area of Real Algebraic Geometry. The physical questions associated
with compatibility of partial information about quantum states may
be encoded in two polynomials in real variables; one polynomial is
quadratic (defining a sphere) and the other cubic. Computational
Real Algebraic Geometry is an active field of current research
with considerable effort being devoted to finding algorithms for
precisely the problems important here. We believe that it is
intrinsically interesting to identify the part of mathematics
within which questions of compatibility of partial information
about quantum states lie. One would also like to have methods for
solving instances of the problems. In fact our problems, perhaps
not surprisingly, are hard in a strict algorithmic sense.
Unfortunately the best algorithms that we are aware of are not
sufficiently powerful at present to be
 useful for interesting cases on modest computers.
Nonetheless we anticipate that, with increased algorithmic and
computational power, this will change. In any case, we argue that
techniques, analytic or algorithmic, from Real Algebraic Geometry
are required to solve problems concerning parts of quantum states.

We start by showing that necessary and sufficient conditions for
an Hermitian matrix, $H$, to be the density matrix of a pure state
(i.e. to be a positive, Hermitian matrix with trace one and
exactly one non-zero eigenvalue) are that,
\begin{equation} Tr
(H^2)=Tr (H^3)=1.\label{eta}
\end{equation}
(Note that $Tr (H^2)=Tr (H)=1$ are not sufficient conditions since
they are satisfied by the non-positive matrix
$H=diag(\frac{1}{2},\frac{1}{2},\frac{1}{2},-\frac{1}{2})$). It is
obvious that the density matrix of a pure state satisfies
(\ref{eta}). To see that (\ref{eta}) are sufficient for $H$ to be
a density matrix, consider a basis for $H$ in which it is diagonal
with diagonal elements $(\lambda_1,...,\lambda_n)$. Then
$\sum_{i=1}^n \lambda_i^2=1$ implies that each $\lambda_i$
satisfies $-1\leq \lambda_i\leq 1$. Then:
\begin{equation}
\sum_{i=1}^n \lambda_i^3\leq\sum_{i=1}^n \lambda_i^2,
\label{lambda_cubed_less_than_lambda_squared}
\end{equation}
with equality only when each  $\lambda_i=0$ or $+1$. But our
conditions (\ref{eta}) require equality in
(\ref{lambda_cubed_less_than_lambda_squared}) and hence to be
compatible with $\sum_{i=1}^n \lambda_i^2=1$, exactly one
$\lambda_i$ must be equal to $+1$.

In the first instance, we will illustrate the general issues
concerning partial data for quantum states by considering the case
of states of three qubits. In this setting it is helpful to write
the state in terms of the Bloch Decomposition. Any Hermitian,
trace one, $8\times 8$ matrix may be written as:
\begin{eqnarray}\rho_{ABC} = & &{\frac 1 8}\Big( 1 \otimes 1\otimes
1 +\alpha_i \sigma_i \otimes 1\otimes 1 + \beta_i 1 \otimes
\sigma_i \otimes 1+ \gamma_i 1 \otimes 1 \otimes \sigma_i + R_{ij}
\sigma_i \otimes \sigma_j \otimes 1 \nonumber\\
& &+ S_{ij} \sigma_i \otimes 1\otimes \sigma_j + T_{ij} 1\otimes
\sigma_i \otimes \sigma_j  + Q_{ijk} \sigma_i \otimes \sigma_j
\otimes \sigma_k \Big),\label{star}
\end{eqnarray}
 where the $\sigma_i$'s are the Pauli matrices (which are Hermitian and traceless) and `$1$' is the $2\times 2$ identity matrix. $(1,\sigma_x,\sigma_y,\sigma_z)$ form a
basis for all $2\times 2$ matrices. Note that
$\alpha_i=Tr(\rho_A\sigma_i),
R_{ij}=Tr(\rho_{AB}\sigma_i\otimes\sigma_j)$ etc. Using this
parametrization of $\rho^{ABC}$, (\ref{eta}) becomes
\begin{eqnarray}
Tr\rho^2&&=\frac{1}{8}(1 + \alpha_i \alpha_i+\beta_i\beta_i+
\gamma_i\gamma_i+S_{ij}S_{ij}+R_{ij}R_{ij}+T_{ij}T_{ij}+Q_{ijk}Q_{ijk})=1,\label{r}\\
Tr\rho^3&&=\;\;\frac{1}{64}\Big[1+3(\alpha_i
\alpha_i+\beta_i\beta_i+
\gamma_i\gamma_i+S_{ij}S_{ij}+R_{ij}R_{ij}+T_{ij}T_{ij}+Q_{ijk}Q_{ijk})+6(R_{ij}\alpha_i\beta_j\nonumber\\&&+\;S_{ij}\alpha_i\gamma_j\;+\;T_{ij}\beta_i\gamma_j\;+\;Q_{ijk}\alpha_iT_{jk}\;+\;Q_{ijk}\beta_jS_{ik}\;+\;
Q_{ijk}\gamma_kR_{ij}\;+\; R_{ki}T_{ij}S_{kj})\nonumber\\&&-6
(R_{ij}R_{kl}R_{mn}+S_{ij}S_{kl}S_{mn}+T_{ij}T_{kl}T_{mn})\epsilon_{ikm}\epsilon_{jln}
-6Q_{ijk}Q_{nop}(R_{lm}\epsilon_{inl}\epsilon_{jom}\delta_{kp}\nonumber\\&&+S_{lm}\epsilon_{inl}\epsilon_{kpm}\delta_{jo}+T_{lm}\epsilon_{jol}\epsilon_{kpm}\delta_{in})\Big]=1.\label{s}
\end{eqnarray}
Note that equation (\ref{r}) defines a sphere. If $\rho$ satisfies
these conditions it is a pure quantum state.

\subsection{Testing a Given set of Parts}\label{given}
At the beginning of the last section we described two types of
questions concerning parts of quantum states. The first was: given
a particular set of partial information about a putative state,
can one determine whether this information is legitimate. In the
case of three qubits we might be given, for example, three
two-party density matrices $\sigma^{AB},\tau^{BC},\eta^{AC}$.
There are some simple compatibility relations which are easily
checked, namely we must have that the one-party states derived
from these two-party states are consistent i.e.
\begin{equation}
Tr_B\sigma^{AB}=Tr_C\eta^{AC};\quad
Tr_B\tau^{BC}=Tr_A\eta^{AC};\quad
Tr_A\sigma^{AB}=Tr_C\tau^{BC}.\end{equation} If these conditions
are satisfied, specifying $\sigma^{AB},\tau^{BC},\eta^{AC}$ is
equivalent to specifying $\alpha_i$, $\beta_i$, $\gamma_i$,
$R_{ij}$, $S_{ij}$, $T_{ij}$ for all $i,j$. Substituting these values
into (\ref{r},\ref{s}) leads to two polynomials in the $27$
variables $Q_{ijk}$ (in fact they are both quadratics in this
case). $\sigma^{AB},\tau^{BC}$ and $\eta^{AC}$ are legitimate
reduced states of some three qubit pure state when there is a set
of $Q_{ijk}$ fulfilling both conditions: i.e. testing
compatibility of particular parts of quantum states is equivalent
to finding real roots of polynomials.

Thus far we have considered the set
$\sigma^{AB},\tau^{BC},\eta^{AC}$ and asked whether they are parts
of a three qubit pure state, but it should be noted that the
structure of the problem is essentially the same (albeit involving
more variables) for much more general situations. More general
partial data could be supplied. For example, one might be given
only $\sigma^{AB}$ and $\tau^{BC}$ or perhaps only $R_{ij}$ and
$\alpha_i$. We might also have asked whether
$\sigma^{AB},\tau^{BC},\eta^{AC}$ are consistent with any mixed
state of $ABC$ (not just a three qubit pure state). One can always
purify a mixed state of three qubits by introducing three further
qubits $DEF$. (The most general state of six qubits may be
expressed in a form analogous to (\ref{star}) but now with terms
up to one of the form
$Z_{ijklmn}\sigma_i\otimes...\otimes\sigma_n$). Thus the question
of the existence of a mixed state compatible with
$\sigma^{AB},\tau^{BC},\eta^{AC}$ is again that of asking if there
is a real solution to a pair of polynomial equations (i.e. the
equations expressing the purity of $\rho^{ABCDEF}$) given some of
the variables. It should be clear that very general questions
concerning compatibility of some particular partial information
about a multi-qubit state can be phrased as whether there are
solutions to a pair of polynomial equations, one quadratic and one
at most cubic in the variables. Finally in this context we note
that for any dimension of local Hilbert space there are bases
consisting of the identity, together with traceless Hermitian
matrices, thus our description is not restricted to qubits.

While it is useful  to understand the general framework within
which questions of compatibility of reduced states lie,
unfortunately,  equations (\ref{r},\ref{s}) are hard to solve in
practice. Finding the real roots of systems of equations in large
number of variables is a frontier area of computational research
\cite{Rou,Aub,Roy,Vega}. Our pair of polynomials, the quadratic
$p_1=Tr(\rho^2)-1=0$ and the cubic $p_2=Tr(\rho^3)-1=0$, can be
expressed as a single order $6$ polynomial: $p^2_1+p^2_2=0$. One
of the best algorithmic bounds for the number of arithmetic steps
to find a real root is exponential in the number of unknowns
\cite{Reneg92} (see also \cite{Blum,BSS89}), and so doubly
exponential in the number of parties.

\subsection{Finding Compatibility Conditions}
In the introductory remarks to this section we described a
slightly different question which we would like to answer. What
conditions must a set of parts of quantum states satisfy such that
there exists a compatible quantum state?

In the case of three qubits, rather than solving for $Q_{ijk}$ for
particular $\alpha_i...T_{ij}$ values (as in Section \ref{given}),
one would rather find the range of $\alpha_i...T_{ij}$ values such
there always exist real $Q_{ijk}$ satisfying (\ref{r},\ref{s}).
Here one solves (\ref{r},\ref{s}) for general $\alpha_i...T_{ij}$,
eliminating $Q_{ijk}$, with the constraint that the variables are
reals. Finding rules that parts of quantum states must satisfy
such there exists a whole is equivalent to solving a system of
equations, in real variables, for a subset of those variables.

Both analytically and algorithmically this is a hard task. The
first implementable algorithm for eliminating real variables from
systems of equations was Collins's \cite{C75} `Cylindrical
Algebraic Decomposition'. In terms of the number of arithmetical
steps required, this has a worst-case running time doubly
exponential in the number of variables (here the number of
variables is the number of real coefficients in our decomposition,
$\sim d^N$ for $N$ $d$-dimensional subsystems). Subsequent authors
(e.g. \cite{BPR96}) have produced theoretical algorithms that are
singly exponential in the number of variables. It should be noted
that this approach generalizes to  $N$ parties, in higher
dimensions and for different sets of parts of interest.

Using a more advanced version of Collins's algorithm \cite{Hoon}
we made a proof of principle in the simplest case. The conditions
for two one-party reduced states $\rho_A$ and $\rho_B$ to be
compatible with a two-party pure state are known: the eigenvalues
of $\rho_A$ and $\rho_B$ must be the same. This is equivalent to
the conditions $\alpha_i=\pm\beta_i,\;\alpha^2_i\leq 1$. The
algorithm showed that these are sufficient conditions by
elimination of $R_{ij}$ from simplified two party versions of
(\ref{r},\ref{s}) \cite{time} (showing necessity was not tractable
during the time allocated). Despite the unfavorable complexity of
 finding compatibility conditions algorithmically,
 there is hope that, with optimized programs, advances can be made.
Known necessary conditions \cite{Han,No} could be shown to be
sufficient for small numbers of parties.

Given these algorithmic challenges, it is natural to apply ad hoc
techniques to solve specific sub-problems. A restricted form of
the above question is whether a set of parts have a compatible
\it{pure} \rm state. There are two conditions which are necessary
but, in general, not sufficient for a set of reduced states to be
parts of a pure quantum states. (1) As noted above, if the  states
have parties in common, their reduced states in these common
parties alone must be the same. E.g. for $\sigma_{AB}$ and
$\tau_{BC}$, $Tr_A\sigma_{AB}=Tr_C\tau_{BC}$. (2) By the Schmidt
decomposition, pairs of reduced states which are a bipartite
division of pure states must have the same eigenvalues. Satisfying
property (1) alone is insufficient for there to exist a compatible
pure state: the set $\sigma^{AB}=\tau^{BC}=\eta^{AC}=1/2(\ket{00}
+ \ket{11})(\bra{00}+\bra{11})$
 have consistent one party reduced states but
there is no pure three party state with these reduced
states~\cite{LPW}. Both (1) and (2) are still not sufficient for a
set of states to be compatible  with a pure state. Consider the
states:
\begin{eqnarray}
&&\sigma_{AB}=\frac{1}{4}\Big[\ket{00}\bra{00}+\ket{11}\bra{11}+\ket{01}\bra{01}+\ket{10}\bra{10}\nonumber\\
&&\;\;\;\;\;\;\;\;\;\;+\ket{00}\bra{01}-\ket{10}\bra{11}+\ket{01}\bra{00}-\ket{11}\bra{10}\Big],\\
&&\tau_{BC}=\frac{1}{2}\Big[\ket{00}\bra{00}+\ket{11}\bra{11}\Big],\\
&&\eta_{AC}=\frac{1}{2}\Big[\ket{00}\bra{00}+\ket{11}\bra{11}\Big].
\end{eqnarray}
These satisfy both properties (1) and (2) but there is no state
compatible with $\sigma_{AB},\tau_{BC},\eta_{AC}$ (the most
general pure state compatible with $\tau_{BC}$ and $\eta_{AC}$ is
$\frac{1}{\sqrt{2}}(\ket{000} +e^{i\theta}\ket{111})$ but this
cannot reduce to $\sigma_{AB}$).

 Diosi \cite{diosi}
notes that, by the Schmidt decomposition, given any pair of
overlapping reduced states of a pure state $\ket{\phi}_{ABC}$
(e.g. $\rho_{AB}, \rho_{BC}$ where systems $A,B,C$ might not be
equidimensional) the original state is almost always the only \it
pure \rm state compatible with them \cite{Nov}. We note that it is
fairly straightforward to convert this result into a set of
compatibility conditions which are necessary and sufficient for
reduced states $\rho_{AB}, \rho_{BC}$ to be the reduced states of
a pure state.

\section{Conclusion}
We have provided tight bounds on the size of the parts that are a
sufficient description of almost all pure states and we have
placed these questions in the framework of Real Algebraic Geometry
which is suitable for all such tasks involving parts of quantum
states.

The first part of this paper tells us that almost all pure quantum
states contain no higher order correlations which are not
determined by correlations within parts slightly bigger than half
the whole state. Exceptional states which do not have this
property are of interest. The $N$-party GHZ,
$\frac{1}{\sqrt{2}}[\ket{0}^{\otimes N} +
e^{i\theta}\ket{1}^{\otimes
 N}]$, has irreducible $N$-party entanglement: its entanglement
 cannot be asymptotically and reversibly converted into entanglement between
 fewer
 than $N$-parties  \cite{BPRST99,LPSW99}. It is also `part-wise' irreducible - no set of
 its reduced states can uniquely determine the state. A better
 understanding of the connection between these two kinds of
 irreducibility, and whether one implies the other, would be
 desirable. It is easy to construct non-generic $N$-party states
 which are not determined by their $m$-party reduced states: it is interesting to understand what
 properties  these states have, and  to find the
 set of all such states (this was done for three qubit states in \cite{LPW}). % A more technical question is whether there
 %are pure states with reduced states with two properties (1) they
% are the reduced states of no other pure state and (2) they are the
% reduced states of a mixed state. Can the algebraic formulation
% provide further insight into these concerns? Is there a
% geometrical feature of the equations that should lead us to
% expect most pure states to be uniquely determined by their $(N/2
% +1)$-party reduced states?
 Finally, while we have
given \em{proofs} \rm that reduced states of roughly half the
number of parties determine $N$-party pure states, we do not have
a simple understanding of why this should be the case.

\section*{Acknowledgements} We are grateful to Fabrice Rouillier, Stanly
Steinberg, Tobias Osborne, Andreas Winter and William Wootters for
helpful discussions. We are grateful for support from the EU, via
the project RESQ.

\end{document}